# Infusing Linguistic Knowledge of SMILES into Chemical Language Models

Ingoo Lee, and Hojung Nam*

School of Electrical Engineering and Computer Science, Gwangju Institute of Science and Technology, 123 Cheomdangwagi-ro, Buk-ku, Gwangju, 61005, Republic of Korea

*: corresponding author

E-mail addresses:

Ingoo Lee: dlsrnsladlek@gist.ac.kr

Hojung Nam: hjnam@gist.ac.kr

## Abstract

The simplified molecular-input line-entry system (SMILES) is the most popular representation of chemical compounds. Therefore, many SMILES-based molecular property prediction models have been developed. In particular, transformer-based models show promising performance because the model utilizes a massive chemical dataset for self-supervised learning. However, there is no transformer-based model to overcome the inherent limitations of SMILES, which result from the generation process of SMILES. In this study, we grammatically parsed SMILES to obtain connectivity between substructures and their type, which is called the grammatical knowledge of SMILES. First, we pretrained the transformers with substructural tokens, which were parsed from SMILES. Then, we used the training strategy "same compound model" to better understand SMILES grammar. In addition, we injected knowledge of connectivity and type into the transformer with knowledge adapters. As a result, our representation model outperformed previous compound representations for the prediction of molecular properties. Finally, we analyzed the attention of the transformer model and adapters, demonstrating that the proposed model understands the grammar of SMILES.

## Introduction

Accurate prediction of molecular properties is critical for the efficient and successful development of drugs. Therefore, researchers have developed chemical models to predict molecular properties. For example, quantitative structure-activity relationship (QSAR) builds a mathematical model that determines the structure's quantitative contribution to molecular properties. Recently, many deep learning models have been developed to predict molecular properties, outperforming previous QSAR methods [1, 2]. For the deep learning model, the input representation of chemical compounds is the most important factor for model architecture and performance. Input representations of chemical compounds can be molecular descriptors [3], classical chemical fingerprints [2, 4], a graph consisting of atoms and bonds [5, 6], and a string notation of chemical compounds [7-9]. For the string notation of chemical compounds, the simplified molecular-input line-entry system (SMILES) is the most popular representation system for chemical compounds. A SMILES representation is generated by traversing molecular graphs using a depth-first search (DFS) strategy. Therefore, SMILES can be converted into an original graph. Because of its simplicity and validity, most chemical databases use the SMILES representation for compounds [10, 11]. Given the abundance of SMILES as simple string form, which are beneficial for deep learning models, many SMILES-based deep learning models have been developed. For example, RNN-based seq2seq models are proposed as potenital chemical fingerprints, which encode and decode to reconstruct their own SMILES [12, 13]. Recently, milestone models for natural language processing (NLP), transformers [14], and Bidirectional Encoder Representations from Transformers (BERT) [15] have been applied to SMILES [7, 8, 16-18]. The attention of the transformer and BERT is a key mechanism for learning the interdependencies between atoms by adjusting the distance of their embedding in the chemical space. In particular, BERT suggested an enhanced strategy to learn interdependencies between words by predicting masked words, which is called the masked language model (MLM). Although SMILES is widely used, it has some inherent limitations. For example, because SMILES traverses molecular graphs by DFS, rings in compounds are broken and connected atoms are distantly located [19]. Similarly, branches of compounds are inserted in SMILES with parentheses, inducing the distance between connected atoms. However, despite its limitations, the transformer and BERT-based models did not validate the connectivity of atoms

[8, 17, 18]. Recent studies of BERT, called BERTology, suggested enhanced modeling and analysis of attention to comprehend dependency [20, 21]. Additionally, the previous model did not build substructural tokens [8, 16], whereas the substructural tokenization of SMILES was validated to increase the prediction performance [22, 23]. Substructural tokens are useful by themselves because they can recognize frequent structural motifs of a compound by itself, without parameter training, gaining a solid ground to understand the structural composition of the chemical compound. As in SMILES, the relationship between tokens is not linear, but hierarchical as in natural language. This can be regarded as the grammatical system of SMILES, with substructural tokens as words and syntactic relationship as the connectivity between atoms contained in tokens. However, previous studies did not explicitly model or analyze the "grammatical" relationship between tokens [18]. In this study, we first trained the BERT model with substructural tokens, which were designed to comprehend the grammar of SMILES. In addition, by training the BERT model to determine whether two compounds are the same, BERT was trained to to understand structural correspondence. After pretraining, we injected "grammatical knowledge" of SMILES by hiring a K-adapter [24], without the catastrophic forgetting of the enriched information from SMILES. Therefore, SMILES can be represented as a tree of substructural tokens, and we validated BERT-like structural correspondence learning and infuse grammatical knowledge to increase molecular property performance. Our model is primarily implemented using the Hugging Face interface [25]. The overall pipeline is shown in Fig. 1 (a).

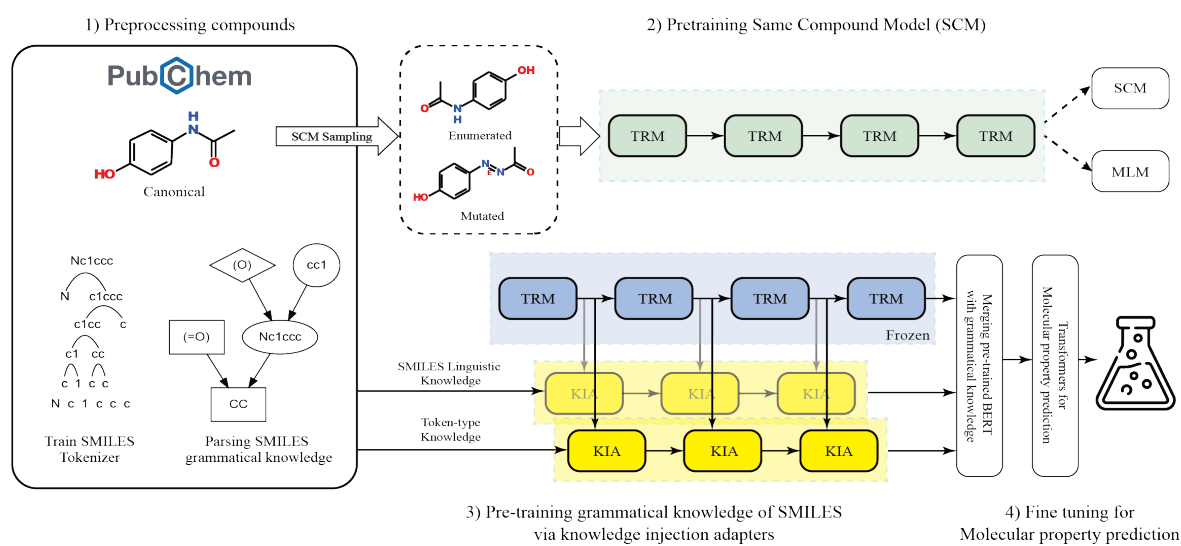

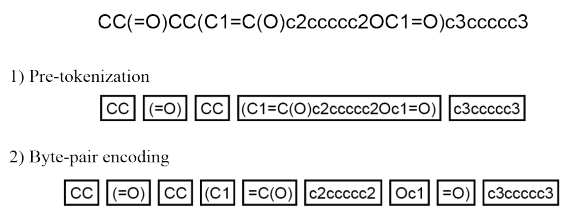

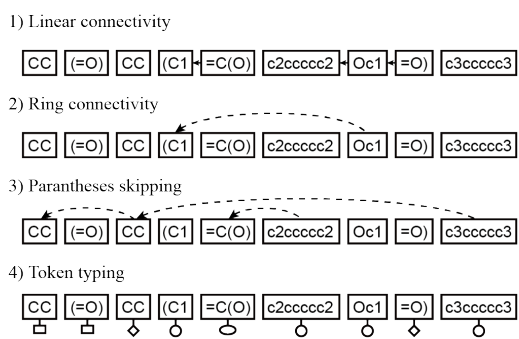

Fig. 1. Model overview. (a) Model overview. Our model uses a four-step process. First, SMILES is preprocessed as sets of tokens and their corresponding knowledges. Second, BERT is pretrained, called same compound model to predict masked tokens and whether next compound is enumerated or mutated. Third, the parsed knowledge is pretrained with K-adapters (knowledge injection adapter, KIA) for enhanced understanding of SMILES for the pretrained BERT. Finally, the molecular properties are predicted using the pretrained BERT and knowledges. (b) Tokenization. Tokenization was processed by two steps, pretokenization by brackets and byte pair encoding. (c) Grammatical parsing of SMILES. We parsed grammatical structure of tokens by their connectivities and types

# Materials and Methods

## Pretraining BERT

### Tokenization of SMILES

Many studies have demonstrated that the tokenization of SMILES increases the SMILES-based prediction performance [22, 23]. Data-driven tokenization of SMILES can detect substructures, including functional groups or motifs that commonly appear as drug-like compounds, without parameter training. However, for better tokenization of SMILES, its grammar needs to be pretokenized. In NLP, BERT first pretokenizes white spaces to recognize words or compound words. However, in SMILES, there is no white space inside. In SMILES, parentheses play the role of grammatical punctuation by indicating the start and end of a branch. Additionally, square brackets are used to indicate atoms and charged atoms. Therefore, we pretokenized SMILES using outer parentheses and square brackets [23] to help the tokenizers recognize the substructures correctly, as shown in Fig. 1 (b). After pretokenization, we trained byte pair encoding [26] to query common SMILES tokens that specify the substructure from PubChem 10M compounds [11], which was gathered from ChemBERTa [18]. We note that all SMILES is non-isomeric to concentrate on further grammatical parsing. Therefore, the SMILES of compounds are represented as a list of substructural tokens.

### Pretraining same compound model (SCM)

We trained the BERT using SMILES. Many previous models employed only MLM to train SMILES on BERT [8] or BERT-based models [18] because it is difficult to apply the next sentence prediction (NSP) model for SMILES. Conversely, many studies have shown that training SMILES with the same but enumerated [7, 27, 28] or slightly mutated SMILES [17] could help increase the sensitivity of detection of molecules' substructures. Therefore, for sensitive recognition of substructures and increased prediction performance, two SMILES were passed to the BERT model to distinguish whether the second SMILES is simply enumerated or slightly modified (mutated), which is called "the $\underline{S}$ame $\underline{C}$ompound $\underline{M}$odel" (SCM). During the BERT training, we randomly sampled the pairing compounds. Half of them was enumerated SMILES but the same and mutated SMILES in the other case. We generated a valid mutation of SMILES with the previous SMILES mutation function [29, 30] that is used in genetic algorithms. By comparing the two SMILES, we expected the model to learn structural correspondences between canonical and enumerated SMILES and recognize a small change in structure, leveraging the understanding of SMILES grammar. We can formulate the SCM as:

$$< CLS > \quad Canon(Mol) \quad < SEP > \quad Enum(Mol) \quad \rightarrow \quad 1$$
$$< CLS > \quad Canon(Mol) \quad < SEP > \quad Mutate(Mol) \quad \rightarrow \quad 0$$

where, <CLS> is the class token and <SEP> is the separation token. We use binary cross entropy as loss for the SCM. Therefore, loss for pretraining chemical BERT model with the SCM will be formulated as

$$L_{BERT} = L_{MLM} + L_{SCM}$$

where $L_{MLM}$ is a loss for the MLM and $L_{SCM}$ is a loss for the SCM. We trained the BERT model with SCM and MLM with PubChem 10M compounds [11, 18] for 15 epochs. The BERT model contains eight transformers with 256 hidden dimensions and eight attention heads.

## SMILES knowledge injection

### SMILES syntax dataset generation

In the SMILES representation procedure, the SMILES generator traverses atoms through their bonds using a DFS strategy. However, the rings in a molecule need to be broken for DFS. The first and last atoms of a ring traversed by the generator were labbeled as ring open and close with number [19]. Therefore, in SMILES, the atoms of ring open and close must be distant even though they are connected in a compound. Additionally, during the traversal of atoms of a molecule, the generator encounters a number of branches. Branches are labelled as enclosures in parentheses, and they may be nested and stacked. Thus, the atom preceding the rounded opening bracket and the one following the rounded closing bracket are connected in a molecule. In summary, the tokens within parentheses should be skipped. It is difficult for the model to capture the connectivities of ring open–close and parentheses skipping. In extreme cases, they can be located at the start and end of the SMILES. First, to train connectivity between atoms, we parsed connectivities between tokens by enumerating SMILES and stacks of ring numbers and parentheses. Tokens are linearly connected as the next token indicates the previous token. In the case of ring open–close and parentheses, the latter token indicates former tokens based on their connectivity, as depicted in Fig. 1 (c).

### Token-type dataset generation

In SMILES, the same tokens can play different roles in a compound composition. For example, the token "2CCCC" can refer to a ring component or a carbon side chain attached to the ring. For better comprehension of the chemical structure, the model should understand the role of tokens in the SMILES composition. Therefore, by querying ring information of atoms in a token, we classified tokens into five type: "is ring", "has ring", "attached to ring", "non-ring", and "non-atom" as seen in Fig. 1 (c).

### Knowledge-injection adapter

BERT-based models fine-tune pretrained BERT models for specific tasks. They can be used in cheminformatics used to predict molecular properties such as brain–blood barrier permeability (BBBP), toxicity, and bioactivities. However, during fine-tuning, the weights of the pretrained model change owing to the loss of a fine-tuning task, which may lead to loss of enriched information of the pretrained model, such as substructure sensitivity. Additionally, the size of trainable parameters for the BERT model is large, whereas the size of the fine-tuning dataset is much smaller than that of the pretraining corpus. To address the aforementioned problems, adapters can be attached to a pretrained transformer model. As the pretrained BERT freezes, adapters with small sizes of parameters are trained [31, 32]. Additionally, K-adapter [24] trained adapters with factual and linguistic knowledge enhance the performance of fine-tuning tasks. The K-adapter contains projection layers and transformers, which take intermediate outputs of the transformer from the pretrained BERT and previous adapter results. Finally, the output features from the pretrained model and adapter are concatenated to predict the linguistic and factual knowledge. We trained K-adapters with SMILES SYNtax (SYN) and Token-Type (TT) knowledge.

### Knowledge training

Before fine tuning, we trained K-adapters with the parsed knowledge to increase the prediction performance. In SYN knowledge training, each token predicts the father index, which indicates an index of the substructural token connected in the molecular graph and located before it in SMILES. The first token is forced to indicate <CLS> token as we expect the <CLS> token to become more representative. For TT knowledge training, each token predicts its own type. We inserted two adapters after the embedding layer and fourth and eighth transformers and those adapters are trained by cross-entropy of knowledge prediction loss with a 1e-5 learning rate. Each K-adapter had 256 hidden dimensions and eight attention heads.

## Molecular property Prediction

### Molecular property datasets

MoleculeNet [32] was used as the benchmark dataset to train and validate the performance of molecular property prediction. Particularly, we choose the BBBP, clinical toxicity (ClinTox), activity to HIV protein, and p53 stress-response pathway activation in Tox21. Additionally, we used scaffold splitting from DeepChem [33] for further performance validation. In scaffold splitting, we fine-tuned the model with 80% of the dataset with the smallest scaffolds and validated 10% of the dataset with larger scaffolds to select hyperparameters. The remaining 10% of the dataset with the largest scaffolds was used to evaluate the test performance.

### Molecular property prediction models

After training the SMILES knowledge on K-adapters, the output features of K-adapters were added or concatenated to those of the pretrained BERT. We inputted this knowledge-comprehensive output to additional transformers that selectively gathered features of the token to predict molecular properties. The features passing all transformers were pooled to predict the molecular properties. Instead of only using the <CLS> token to predict, we followed the pooling strategy of ProtTrans [34], which utilizes the <CLS> token, global average pooling, global max pooling, and global pooling averaging with squared root. Finally, we placed dense layers to predict molecular properties. While fine tuning, the pretrained BERT and adapters were frozen to preserve their enriched knowledge. We fine tuned 15 epochs with a 1-e5 learning rate with the training dataset, and the best model was selected by the loss of the validation dataset.

# Results and discussion

## Knowledge training results

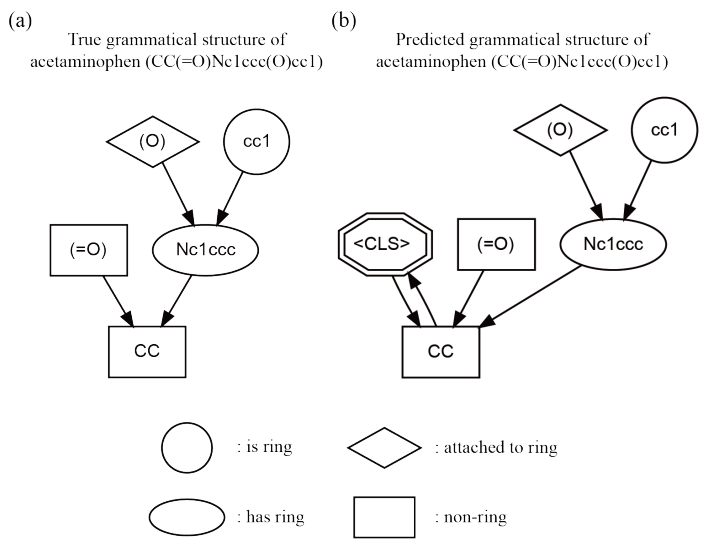

Fig. 2. (a) Syntactic structure of acetaminophen. (b) Grammatical prediction results of SMILES, whose first token was trained to indicate <CLS> token.

From parsed SN and TT knowledge, we could build a grammatical tree of SMILES. For example, as shown in Fig. 2 (a), acetaminophen can be represented as a grammatical tree with its substructural tokens and connectivities. The first token of acetaminophen plays the role of the root for a grammatical tree, and the other tokens indicate their parent token by the connectivity of atoms within. Additionally, the type of tokens is represented by different node shapes: circle, eclipse, diamond, rectangle, and octagon, respectively, for "is ring", "has ring", "attached to ring", "non-ring", and "non-atom". Using SMILES syntax and TT adapters, we built a grammatical tree with SMILES tokens, as shown in Fig. 2 (b). It can be seen that the predicted grammatical tree of acetaminophen is the same as the ground truth. In grammatical tree of SMILES, a first token is used as root of tree, the other tokens eventually indicate a first token. Because we trained the first token to indicate a <CLS> token in knowledge fine-tuning, the first token of acetaminophen referred to the <CLS> token. Although we did not train the <CLS> token to indicate any specific token in knowledge fine-tuning, the <CLS> token mutually refers to the first token. Thus, we expect the <CLS> tokens of SMILES to be represent compounds after constructing a grammatical tree of SMILES.

## Molecular Property Prediction Results

**Table 1. Molecular Prediction Performances**

| Method \ Dataset | BBBP | | ClinTox | | HIV | | Tox21 | |
|---|---|---|---|---|---|---|---|---|
| | ROC | PRC | ROC | PRC | ROC | PRC | ROC | PRC |
| RF | 0.681 | 0.692 | 0.693 | 0.968 | 0.780 | 0.383 | 0.724 | 0.335 |
| SVM | 0.702 | 0.724 | 0.833 | 0.986 | 0.763 | 0.364 | 0.708 | 0.345 |
| Transformer-CNN | 0.820 | 0.697 | 0.841 | 0.615 | 0.792 | 0.287 | 0.416 | 0.219 |
| D-MPNN | 0.708 | 0.697 | 0.906 | 0.993 | 0.752 | 0.152 | 0.688 | 0.429 |
| ChemBERTa 10M | 0.643 | 0.620 | 0.733 | 0.975 | 0.622 | 0.119 | 0.728 | 0.207 |
| SCM | 0.770 | 0.640 | 0.981 | 0.906 | 0.713 | 0.245 | 0.631 | 0.337 |
| SCM with SYN | 0.790 | 0.667 | 0.999 | 0.990 | 0.774 | 0.304 | 0.698 | 0.472 |
| SCM with TT | 0.777 | 0.691 | 0.996 | 0.955 | 0.753 | 0.216 | 0.709 | 0.439 |
| SCM with SYN+TT | 0.814 | 0.744 | 0.994 | 0.954 | 0.797 | 0.296 | 0.741 | 0.487 |

**ROC**: Area under receiver operating characteristic curve
**PRC**: Area under precision-recall curve

We evaluated four models for molecular property prediction on subsets of MoleculeNet [35], and the results are summarized in Table 1. First, we fine-tuned the SCM without freezing the transformers. For the rest of the models, we trained knowledge adapters using frozen SCM (SYN, TT, SYN+TT). Compared with ChemBERTa 10M, we can see that training even structural correspondence in SMILES (SCM) increases the molecular property prediction performance. While RoBERTa [36] shows robust performance compared to BERT [15], we validated that the performance increase by randomizing SMILES is a more effective approach to SMILES [7, 27, 28]. Additionally, we verified that the training knowledge of SMILES increases molecular prediction performance. The proposed model with SYN and TT adapters outperformed previous methods, such as the Transformer-CNN [7] and D-MPNN [37]. In particular, our methods showed a significant performance increase in the toxicity datasets, ClinTox, and Tox21.

## Attention Analysis

### Attention analysis on SCM

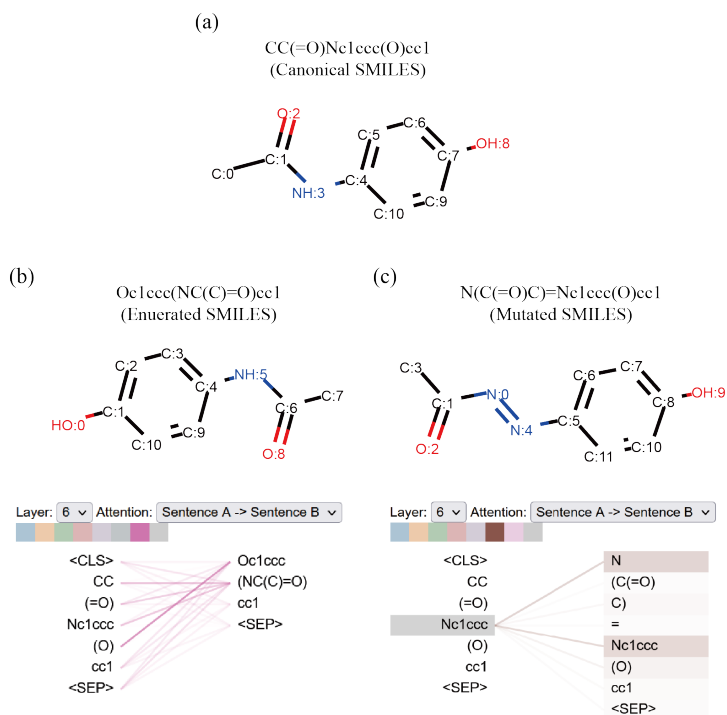

Fig. 3. Attention analysis on SCM. (a) Acetaminophen with canonical traverse numbering. (b) Acetaminophen with enumerated traverse numbering and attention visualization of enumerated acetaminophen on head 7 at transformer 7. (c) Mutated acetaminophen and with enumerated traverse numbering and attention visualization of mutated acetaminophen on head 6 at transformer 7

We inspected the attention of the SCM to understand the mechanism of structural correspondence. Particularly, we selected acetaminophen for the simplicity of attention analysis. First, we randomized SMILES of acetaminophen and input canonical and randomized SMILES to the SCM. We inspected the attention of SCM by BertViz [38]. Interestingly, although the order of the two oxygens in randomized SMILES is flipped from canonical SMILES, at head 7 of transformer 7, we can see that the second oxygen of canonical SMILES indicates the first oxygen of randomized SMILES, which is the same oxygen in acetaminophen, as shown in Fig. 3 (b). Conversely, in the case of mutated SMILES where one more nitrogen is added with a double bond, at head 6 of transformer 7, we can see that the nitrogen of canonical SMILES gives attention to both nitrogen atoms of mutated SMILES, as shown in Fig. 3 (c).

## Attention analysis on the knowledge adapters

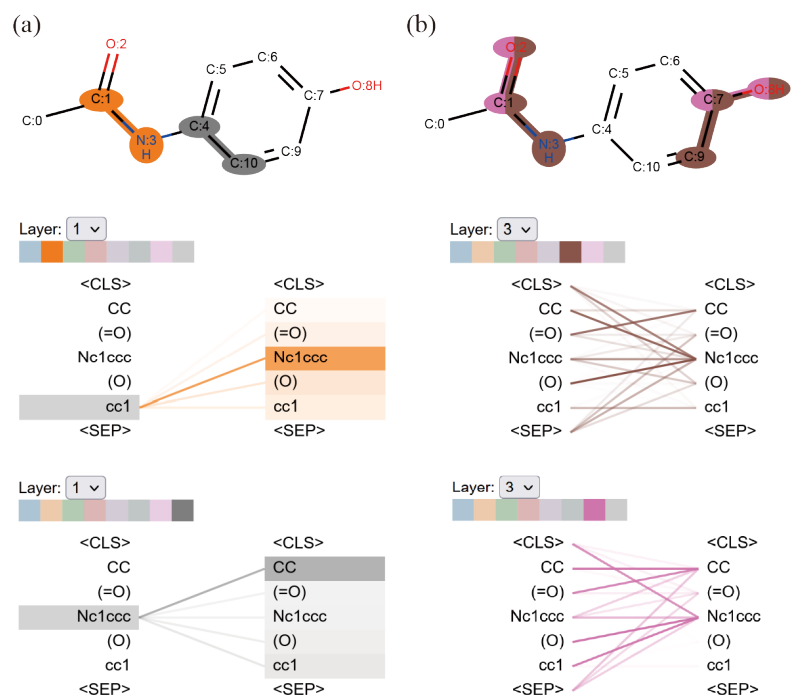

Fig. 4. Attention analysis on adapters. (a) Top: Expected connectivies of atoms by attention of SN adapter. Middle: attention visualization at head 2 of transformer 2. Bottom: attention visualization at head 8 of transformer 2. (b). Top: expected connectivities of atoms by attention of TT adapter. Middle: attention visualization at head 6 of transformer 4. Bottom: attention visualization at head 7 of transformer 4.

We first inspected the attention of the SYN adapter. Interestingly, at the second transformer, some heads inferred connectivities from ring open–close and parenthesis skipping (Fig. 4 (a)). For example, token "cc1", containing a ring close to head 2, clearly assigned high attention to its pairing ring open. Additionally, head 8 inferred parenthesis skipping. The TT adapter, although not trained using the SMILES syntax knowledge, assigned attention to connectivities between tokens but not as clearly as the SYN adapter. For example, as shown in Fig. 4 (b), at head 6 of transformer 4, the first token "CC" indicates the ring component "Nc1ccc" skipping the oxygen of the carbonyl group. At head 7 of transformer 4, tokens of carboxyl groups give high attention to themselves.

## Limitations

Although we enhanced the BERT model for SMILES by injecting grammatical knowledge of the SMILES, there are several limitations to our model. First, because we assumed linear connectivity of tokens, the first token is the root of a syntactic tree, regardless of whether it is the most representative substructure. Second, we used very simple types of tokens based on ring information. Although simple token typing increases molecular property performance, more elaborate token-typing methodologies such as SMART matching of functional groups, can further increase performance and interpretability of prediction results.

## Conclusions

SMILES is a linear string notation for chemical compounds. Owing to the convenience of SMILES representation, it has become the most popular system to describe molecules and is used as an input for many machine learning models. Recently, many methods have been suggested to better understand SMILES, such as enumeration and SMILES pair encoding. However, these models did not comprehend the inherent "grammar" of SMILES, weakening its representative power. In this study, we built grammatical knowledge of SMILES to help understand it for a model while also adopting previously proposed methods. Specifically, we injected grammatical knowledge of SMILES into the BERT model with a K-adapter without forgetting the pretrained BERT weights. Our results demonstrate that SMILES grammatical knowledge injection increases molecular prediction performance and the model's understanding of the language of SMILES.